# Unusual flat and extended morphology of intercalated Cu under MoS$_2$

Dapeng Jing[1,2], Yong Han[1,3], James W. Evans[1,3], Marek Kolmer[1], Zhe Fei[1,3] and Michael C. Tringides[1,3]

[1] Ames Laboratory, Ames, IA 50011, USA
[2] Materials Analysis and Research Laboratory, Iowa State University, Ames, IA 50011, USA
[3] Department of Physics and Astronomy, Iowa State University, Ames, IA 50011, USA

**Abstract**

A general method was developed to intercalate metals under layered materials through a controlled density of sputtered defects. The method has been already applied to study a range of metals intercalated under graphite and different types of morphologies were realized. In the current work, we extend the method to the study of intercalation under MoS$_2$ noting that work on this system is rather limited. We use Cu as the prototype metal for comparison with Cu intercalation under graphite. Although the growth conditions needed for intercalation under graphite and MoS$_2$ are similar, the type of intercalated phases is very different. Each Cu island which nucleates on top of MoS$_2$ during Cu deposition provides material that is transferred below MoS$_2$, through sputtered defects under the island base; this transfer results in a uniform intercalated Cu "carpet" morphology that extends over the mesoscale. On the contrary, Cu intercalation under graphite results in well separated, compact islands formed by monomer detachment from small Cu islands on top and transfer below through defects far from the islands. The structural techniques (scanning electron microscopy and atomic force microscopy) and spectroscopic techniques (x-ray photoelectron spectroscopy and energy dispersive spectroscopy) are used for the characterization of the intercalated Cu layer.

**I. Introduction**

A major recent effort was to establish subsurface growth of metal islands in graphite, through a controlled density of sputtered defects, as a novel synthesis route to grow nanostructures [1-3]. The randomly deposited atoms diffuse to the defects, move below, and nucleate crystalline islands in the confined spaces of the subsurface layers. The islands are draped by graphene on top. The work has shown a new metal growth mode exhibiting a flatter island morphology compared with islands on top, while the metal is chemically protected by graphene (i.e., the top graphite layer) at ambient conditions. Evidence of graphene draping the islands is seen from STM imaging at the top and island sides, where the graphene lattice is resolved and forms Moiré patterns. The latter results from coincidence between the metal and graphene lattices. For example, 10 graphene unit cells match 9 Ru(0001) unit cells [3]. Several metals have been studied to determine the encapsulated morphologies: Cu [2,4,5], Fe [6,7], Pt [8,], Dy [9,], Ru [3,], Ag [8], Au [8].

One natural question is whether the technique can be employed with other layered materials besides graphite, which have not been studied as extensively for metal intercalation. Preference for intercalation is thought to be a result of the competition between charge transfer (across the metal and the 2D material interface) and strain introduced by the metal confined in subsurface spaces. It is important to explore differences and similarities between intercalation of the same metal on different substrates to understand the factors controlling the intercalation mechanism.

This study focuses on Cu intercalation in $MoS_2$. Previous analysis of Cu growth on top of $MoS_2$ at elevated temperature ~900 K results in multi-height pyramidal islands [10], with bimodal island size distribution. Extensive characterization of the growth of metals on suspended $MoS_2$ has also been performed with TEM [11,12]. In the current experiments deposition of Cu at slightly higher temperature on sputtered $MoS_2$ shows that the well separated islands develop a surrounding flat ring, labeled the "carpet", that grows monotonically with temperature or postdeposition annealing time. These "carpet" regions were studied with several complementary techniques: x-ray photoelectron spectroscopy (XPS) and scanning electron microscopy (SEM) (using either secondary or backscattered electrons) with energy dispersive spectroscopy (EDS) and atomic force microscopy (AFM). These experiments have confirmed that the chemical composition of the "carpet" shows strong Cu signal which is encapsulated by $MoS_2$. The "carpet" thickness is approximately ~5nm and the spatial spreading of its area shows an increasing rate with lapsed time.

For Cu intercalation, similar annealing temperatures are needed for $MoS_2$ and graphite, although substantial intercalation occurs during postdeposition evolution (at the deposition temperature) for $MoS_2$ versus intercalation is observed only during deposition for graphite. A major difference is that encapsulated Cu islands under graphite on average have heights larger than under $MoS_2$; and they are compact, well-separated and with smaller lateral sizes. On $MoS_2$ the intercalated region is planar; it initially surrounds each island and extends laterally during postdeposition evolution to distances comparable to the island separation larger than tens of micrometers. Since the Cu island density on $MoS_2$ is lower by four orders of magnitude than the one on graphite, this also indicates that the kinetic processes involved in the two cases are different. As modelled in ref. [13] for graphite, initially small Cu islands form at defects on top of graphite, but for sufficiently high temperature Cu atoms detach, diffuse on top until they encounter free defects and move below. On the other hand, for Cu on $MoS_2$, fewer and larger islands nucleate on top of defects. With increasing temperature, Cu atoms move through defects at the base of the islands and supply the material for the "carpet" to expand from each one of the islands.

A controlled density of defects, generated not by sputtering, but by plasma treatment [15], has also been used in graphene intercalation for graphene growth on SiC (Gr/SiC). The current experiments and the identification of the key controlling factors in transferring metal below through defects can be relevant in these experiments as well.

## II. Experimental methods for Cu intercalation of $MoS_2$

An Omicron ultrahigh vacuum (UHV) chamber was used for $MoS_2$ preparation with base pressure in the low $10^{-11}$ mbar range, followed by Cu deposition and subsequent XPS characterization. The sample was taken out of the UHV and transferred in air for SEM/EDS, and AFM experiments [1–3]. An Omicron high power resistive heater was used to heat the sample. Sample temperatures were determined using a type of PSC-DG42N infrared pyrometer, with emissivity set to 0.85 [16]. An FEI Quanta FEG 250 field emission microscope was used for SEM imaging to produce images showing topographic and compositional contrasts. To achieve optimal image quality, a 10 keV electron beam was used for imaging. An Oxford low-Z spectrometer with a large area detector (X-Max 80) together with Aztec analysis package was used for EDS analysis. In order to limit the size of the primary beam excitation volume and achieve enhanced surface sensitivity, a 6 keV electron beam was used for the EDS experiments. For XPS, a flood type lab x-ray source (unmonochromated) was used. The Mg anode was chosen for better energy resolution. The photoelectron take-off angle was 45° with respect to the surface normal. The spectrometer was calibrated to give the Au $4f_{7/2}$ binding energy at 84.0 eV and the Cu $2p_{3/2}$ binding energy as 933.0 eV for sputter-cleaned metallic gold and copper surfaces. 100 eV pass energy was used for survey scans with a step size of 1.0 eV; for narrow scans 20 eV pass energy was used with step size 0.1 eV. Under these conditions, measurement of a sputter-cleaned gold film yielded 1.05 eV full-width at half-maximum of the Au $4f_{7/2}$ peak. Analysis of XPS data was carried out using CasaXPS software [17]. AFM images were acquired in tapping mode using a Bruker Dimension Icon scanning probe microscope. All images were acquired in air. Images were post-processed using second-order plane fitting and/or zeroth-order flattening with Nanoscope software.

We followed the preparation of the MoS$_2$ surface and subsequent Cu deposition from a previous publication [10] with one modification: before Cu deposition, the sample was ion bombarded (Vacuum Microengineering, Inc. Model IPS3D) with 1 keV Ar$^+$ ions for 60 s followed by annealing at 900 K for 2 hrs to remove residual Ar. Cu was deposited via physical vapor deposition from an e-beam evaporator. The Cu flux was 11 ± 1 monolayers (ML) of Cu per minute. The MoS$_2$ sample was held at elevated temperatures ($T_{\text{dep}}$) during Cu deposition and for subsequent analysis of postdeposition evolution.

## III. Results

**Cu growth mode and encapsulation as a function of temperature**

When Cu is deposited on pristine MoS$_2$ (*p*-MoS$_2$) and on ion-bombarded MoS$_2$ (*i*-MoS$_2$) at $T_{\text{dep}}$ = 900 K and 950 K (Fig. 1 (a-c)), three-dimensional (3D) Cu islands form on the surface. The majority of these islands appear to be 3-sided pyramids with 0 to 3 corners missing. These 3D islands have been observed for Cu deposition on *p*-MoS$_2$ at a lower temperature of 780 K. These islands were shown to have three Cu(311) side faces with a Cu(111) base [6].

At slightly higher temperature $T_{\text{dep}}$ = 1000 K, a new feature emerges in Fig. 1 coexisting with the 3D islands. As seen in the secondary electron (SE) image (Fig. 1(d)), most of the 3D islands are surrounded by "carpet" areas with a slightly brighter color than the bare MoS$_2$ substrate. SE images show topographic contrast and in addition the SE signal is enhanced by sharp edges in the scanned area. These brighter annular areas indicate higher elevation and therefore are good candidates for intercalated metal areas. They exhibit irregular perimeters. In the following subsections, we present evidence that these higher and flat features are Cu intercalated regions which are covered by a top S-Mo-S trilayer (TL) of MoS$_2$ substrate.

Cu encapsulation by MoS$_2$ is possible as under graphite. However, the emerging morphology is dramatically different since under graphite taller compact Cu islands are encapsulated and well separated. While Fig. 1(d) shows the extended, flatter annuli that seem to spread out from each of the pyramidal islands. The island density in Fig. 1(d) is $1\times10^{-2}$ isl./μm². The island density of encapsulated Cu islands at the same deposition temperature under graphite is ~4 orders of magnitude larger $1\times10^{2}$ isl./μm². This large difference in island density is also consistent with the different growth modes observed, i.e. separate compact islands on graphite grown by material detaching from clusters on top, and moving below through far away defects. On MoS$_2$ the very big islands present are also the ones feeding the spreading annuli through one or several defects underneath their base.

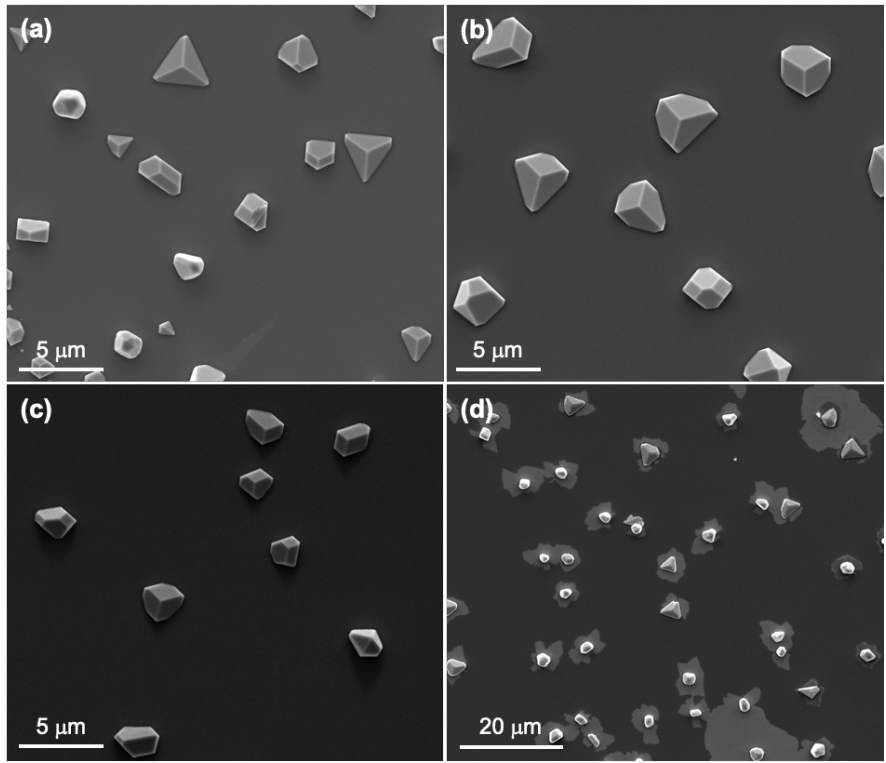

**Figure 1:** SEM images of Cu deposited on *p*-MoS$_2$ surface at (a) $T_{dep}$ = 900 K and on *i*-MoS$_2$ surface at (b) $T_{dep}$ = 900 K; (c) $T_{dep}$ = 950 K and (d) $T_{dep}$ = 1000 K. Only at 1000 K the "carpet" surrounding the islands is seen. As will be discussed next based on spectroscopic analysis the "carpet" is encapsulated Cu and grows from material transport through defects at the island base. With increasing postdeposition annealing time and temperature, the "carpet" spreads laterally while the islands shrink in size.

**EDS analysis of the "carpet" confirming the presence of Cu**

Fig. 2 presents EDS results obtained after Cu deposition on *i*-MoS$_2$ at $T_{dep}$ = 1000 K. Fig. 2(a) shows a high magnification backscattered electron (BSE) image of two 3D Cu clusters and their annuli which have merged together because of cluster proximity. In the

BSE image, the Cu clusters as well as the annuli appear slightly darker than the bare MoS$_2$ substrate. This is the opposite of what is shown in the SE images in Fig. 1. The contrast reversal is due to the fact that BSE images show compositional contrast where material with a lower atomic number appears darker. Cu has a lower atomic number ($Z$ = 29) than the stoichiometry-weighted average atomic number of MoS$_2$ ($Z$ = 31.6). The corresponding Cu $L$ series x-ray map shown in Fig. 2(b) clearly shows the presence of Cu in the annuli. EDS spectra were also collected from three areas A, B and C highlighted in Fig. 2(a) . Spectrum A is from the bare MoS$_2$ and spectra B and C are from the annuli. Spectra B and C show the same Cu $L$ series x-ray intensity at ~0.93 keV indicating the same thickness of Cu in areas B and C. On the other hand, Spectrum A shows no Cu signal above the baseline, thus proving that Cu is only present in the "carpet" and in the tall Cu islands on top. It is also worth noting that no significant O $K$ series x-ray signal was detected in the area shown in Fig. 2(a) including regions A, B and C (data not shown). This indicates the absence of surface oxide formation on the bare Cu cluster, the Cu intercalated carpet region and bare MoS$_2$ substrate after brief exposure to air.

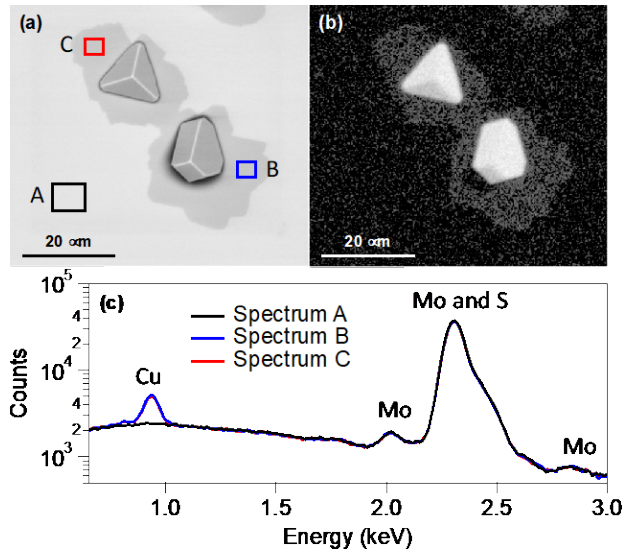

**Figure 2:** (a) BSE image showing the "carpet" and Cu islands. (b) Corresponding EDS map from Cu $L$ series after Cu deposition on $i$-MoS$_2$ at $T_{dep}$ = 1000 K showing Cu in the "carpet" and in the Cu islands. (c) EDS spectra from the three areas highlighted in (a): A (black), on bare substrate, B (blue) and C (red), on "carpet". (Notice the logarithmic scale on the ordinate axis).

To estimate the thickness of the intercalated Cu film, we used the x-ray intensities of the Cu $L$ series in the Cu intercalated regions B and C; as well as the Mo $L$ and the S $K$ series in Fig. 2. (Another method to measure the Cu thickness more precisely is presented in Fig. 3.) We compared the measured intensities with the ones predicted from Monte Carlo simulations using CASINO [17]. A beam of 6 keV electrons was used to excite the sample, and the integrated intensities of the fluoresced Cu $L$ at 0.93 keV and the sum of Mo $L$ and S $K$ x-rays at ~2.3 keV were measured (Mo $L$ and S $K$ signals overlap significantly in EDS).

The integrated areas of the Cu L, Mo L and S K signals are shown in table 1. The intensity ratio of Cu to the sum of Mo and S intensities was experimentally determined as 1:20.4 in region B and 1:19.6 in region C. Subsequently, in CASINO (we defined the sample as a continuous film of Cu encapsulated by a TL of MoS$_2$ on top. We then ran simulations to estimate the intensities of Cu $L$, Mo $L$ and S $K$ x-rays and their ratio that have reached the EDS detector for various Cu film thicknesses and the top MoS$_2$ layer thicknesses, to compare with the ones experimentally determined. We found that 2-3 nm of Cu on MoS$_2$ yielded an intensity ratio closest to the experimentally measured value. The ratio of Cu to the sum of Mo and S is very sensitive to the thickness of Cu and decreases from a value of 60 at ~1nm thickness of intercalated Cu to 16 at ~ 3nm thickness.

Table 1: Experimental measured Cu $L$ series, Mo $L$ series and S $K$ series EDS intensity

|            | Cu $L$ series | Mo $L$ series | S $K$ series | (Mo+S)/Cu ratio |
|------------|---------------|---------------|--------------|-----------------|
| Spectrum A | n/a           | 245787.0      | 234009.0     | n/a             |
| Spectrum B | 22530.0       | 232382.0      | 228411.0     | 20.45           |
| Spectrum C | 23677.0       | 234744.0      | 228510.0     | 19.57           |

In the above CASINO analysis, we assumed the thickness of the encapsulating MoS$_2$ is one TL. This is a reasonable assumption based on the observations from past studies of Cu and of other metal intercalation under graphite [1–3,6,8,9]. Although we note that encapsulation of Cu can occur by up to three layers under graphite [2]. For MoS$_2$, kinetically one expects the Cu atoms to move easier at the TL immediately below the surface (1 TL). This is better justified for Cu moving below MoS$_2$ (than below graphite). For the latter, the graphene layer is only a single carbon atom thick, so transfer under more than one graphene layer is viable. For MoS$_2$, traversing a three-atom (S-Mo-S) thick TL, is kinetically more challenging and involves transfer through a larger number of substrate atoms.

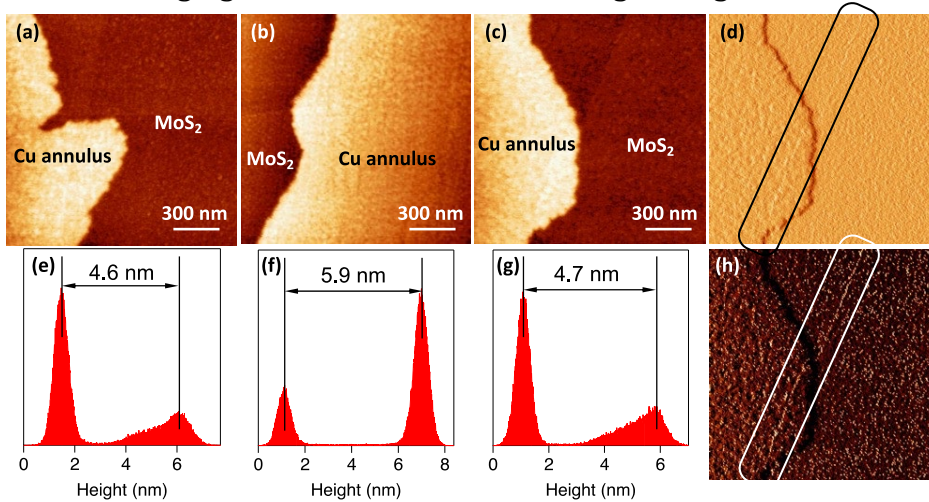

**Figure 3:** (a-c) AFM images of typical boundaries between the intercalated Cu "carpet" and bare MoS$_2$ and (e-g) corresponding height histograms of the previous images showing the height distribution of different "carpet" regions. The widths of the peaks shown in the histograms are caused by the finite tip radius. (d) Derivative mode of images shown in (c); and (h) further equalized images of (d) showing a step edge (circled) of the MoS$_2$ crossing the boundary between intercalated Cu and the bare substrate.

**AFM characterization of the Cu "carpet"**

We employed AFM to directly measure the thickness of the Cu annulus after deposition on i-MoS$_2$ at $T_{\text{dep}}$ = 1000 K. Fig. 3 shows three typical boundaries between the annulus and bare MoS$_2$ and the corresponding height histograms from the AFM images. These images show that the thickness of the Cu film ranges from 4 nm to 6 nm, in good agreement with the previous EDS estimate.

For comparison, in the Cu growth on graphite, we measured the dimensions of the intercalated Cu clusters [2]. The height ranges from 1.5 nm to 43 nm and the diameter from 34 to 607 nm. These encapsulated Cu islands underneath graphite usually exhibit a faceted, quasi-hexagonal footprint. Two types of islands were observed, one with flat and one with round top [2]. In the case of Cu annulus on *i*-MoS$_2$ at $T_{\text{dep}}$ = 1000 K, the height of encapsulated Cu falls on the lower end of the height range for Cu under graphite. However, the lateral extent of these Cu annuli is very different and is much larger than the diameter of the Cu islands under graphite.

In Fig. 3(c), a MoS$_2$ step edge is seen (i.e., a native substrate step separating adjacent MoS$_2$ terraces, not a step in the encapsulated Cu film). This step is hard to resolve because the MoS$_2$ step height 0.7 nm[18] is smaller than the height of the encapsulated Cu "carpet" 4.7 nm (Fig. 3(c)). Derivative (Fig. 3(d)) and further processed (Fig. 3(h)) images are presented to highlight the MoS$_2$ step edge. This MoS$_2$ step intersects the boundary between encapsulated Cu and bare substrate, which indicates that the encapsulated Cu grows without interruption across the step (similar to the Cu encapsulation under graphite[2]).

**XPS evidence of encapsulation**

In this subsection, we show additional evidence that the Cu in the annular region surrounding the 3D Cu clusters is intercalated under MoS$_2$. First, Fig. 4(a) shows the survey spectra after Cu depositions at $T_{dep}$ = 900 K on $p$-MoS$_2$ and $T_{dep}$ = 1000 K on $i$-MoS$_2$. Core level and Auger signals from Cu, in addition to Mo and S are observed. The survey spectra also show the surfaces are oxygen free (O 1s signal appears at ~528-533 eV). Next, the conclusion that Cu is intercalated is supported by the difference in the inelastic tail of the Cu 2$p$ spectra shown in Fig. 4(b). An increase in the slope in the inelastic tail of the Cu 2$p$ region at 1000 K indicates a change in morphology of the deposited Cu [19,20] (see Fig. 4(d)). When Cu is encapsulated into the galleries of MoS$_2$, the ejected Cu 2$p$ photoelectrons experience additional inelastic scattering traversing the MoS$_2$ layers on top (i.e., more kinetic energy is lost). Therefore, Cu 2$p$ intensity shifts to lower kinetic energy (left of the peak, as if these electrons originate from states of higher binding energy). This causes the inelastic tail to rise.

It is expected that some contribution to the inelastic tail of the Cu 2$p$ peak might originate from Cu self-attenuation (see Fig. 4(d)). Such attenuation is possible if there is Cu island coarsening: when the average size of the clusters increases but the density of the clusters decreases. However, coarsening was not observed in this case. The average 3-d cluster size and density are the same for the $T_{dep}$ = 900 K deposition on $p$-MoS$_2$ and the $T_{dep}$ = 1000 K deposition on $i$-MoS$_2$. Therefore, the rise of the inelastic tail due to Cu self-attenuation should be the same in both cases. The rise of the inelastic tail in the 1000 K case can only be attributed to the intercalated Cu in the "carpet" that is only present at 1000 K and not at 900 K. An increase in surface roughness can possibly cause the inelastic tail of the core-level XPS spectrum to rise. However, this possibility can be ruled out by comparing the sample surfaces after Cu deposition on $p$-MoS$_2$ and the $T_{dep}$ = 1000 K deposition on $i$-MoS$_2$. On the macro scale, the scotch tape cleaving method produced intact and flat MoS$_2$ surfaces for all the experiments. SEM imaging on freshly-cleaved pristine MoS$_2$ surfaces (data not shown) show flat regions separated by step bunches. On the micro scale, except for the "carpet" features, Figs. 1(a) and 1(d) show very similar surface morphology of Cu islands on top of MoS$_2$. As the "carpet" spreads and Cu islands decrease in volume there is no increase in surface roughness.

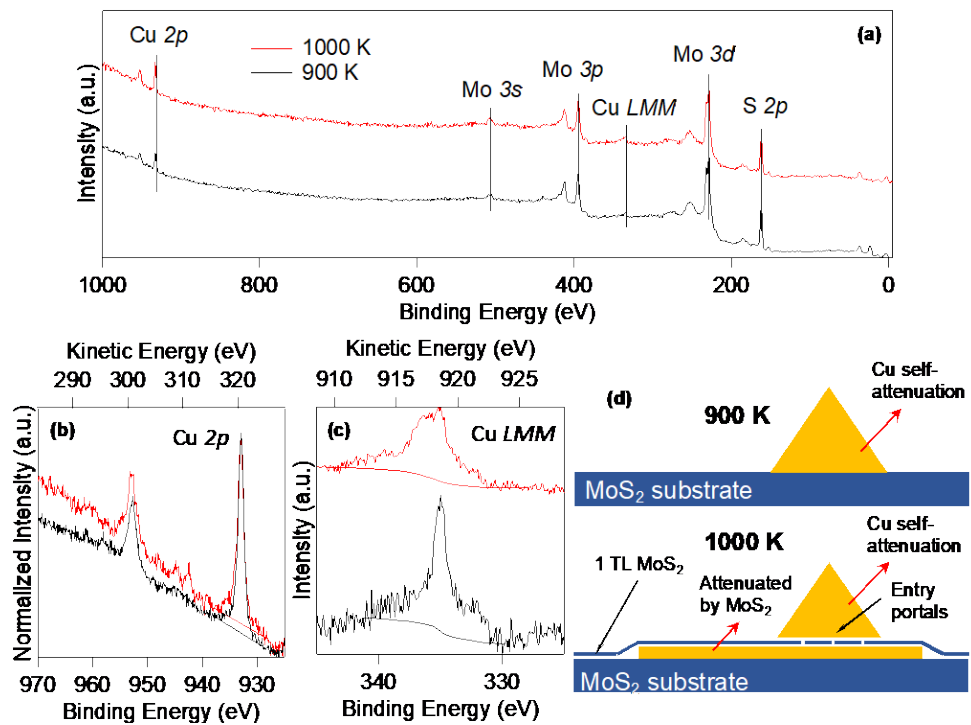

**Figure 4:** XPS spectra of (a) Survey, (b) Cu *2p* and (c) Cu LMM after Cu deposition on p-MoS$_2$ at $T_{dep}$ = 900 K (black) and on i-MoS$_2$ at $T_{dep}$ = 1000 K (red). Notice that the slope of the red curve in (a) is higher than the slope of the black curve which as discussed in the text indicates that the "carpet" is covered by MoS$_2$. (d) Schematics showing the bare Cu cluster after 900 K deposition (top) and the intercalated Cu covered by 1 TL of MoS$_2$ underneath a bare cluster (bottom). Entry portals are shown at the bottom of the bare Cu cluster. The same self-attenuation (at 900 K and 1000K )and attenuation (only at 1000 K) induced by the 1 TL MoS$_2$ of Cu *2p* photoelectrons and Cu LMM Auger electrons are depicted by the red arrows. The size of the bare clusters and the thickness of the intercalated Cu film are not drawn to scale.

Additional evidence for Cu intercalation is seen in the intensity ratio of the Cu 2*p* to the Cu LMM Auger signal. Ejected electrons in these peaks have different kinetic energy. Cu 2*p* core level photoelectrons have a lower kinetic energy of 321 eV (Fig. 4(b)) compared to the energy of the Cu LMM photoelectrons at 918 eV (Fig. 4(c)). The Cu LMM Auger electrons have a larger inelastic mean free path and are attenuated less by material on top. As a result, inside the Cu annulus, inelastic scattering loss for the Cu LMM electrons when passing through the draping MoS$_2$ layer should be less than that for the Cu 2*p* core level photoelectrons. Therefore, if more Cu is encapsulated, even though both Cu 2*p* and Cu LMM intensities decrease due to inelastic scattering, the Cu 2*p* intensity decreases by a larger factor. As a result, the Cu LMM:Cu 2*p* ratio should increase. Indeed, as shown in Table 2, this ratio increases from 0.31 for $T_{dep}$ = 900 K Cu deposition on p-MoS$_2$ (without "carpet") to 0.35 for $T_{dep}$ = 1000 K Cu deposition on i-MoS$_2$ (with encapsulated Cu in the "carpet"). This increase of the ratio is due to part of the Cu is encapsulated whereas all Cu is on top at $T_{dep}$ = 900 K.

**Table 2:** Cu $2p_{3/2}$ and Cu LMM peak intensities from intergrated peak areas using the linear baseline for Cu $2p$ and Shirley baseline for Cu LMM shown in Figs. 4(b) and 4(c).

|  | 900 K | 1000 K |
|---|---|---|
| Cu $2p_{3/2}$ | 1389.6 | 1026.6 |
| Cu LMM | 431.3 | 362.2 |
| Cu LMM:Cu $2p_{3/2}$ | 0.31 | 0.35 |

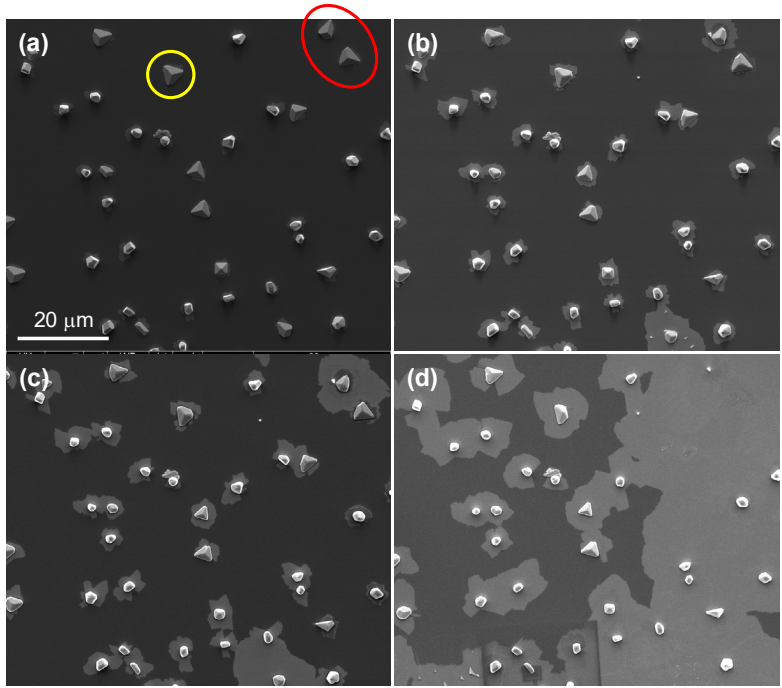

**Figure 5:** SEM images of the same area (a) after Cu deposition on i-MoS$_2$ at $T_{dep}$ = 1000 K; (b) following postdeposition annealing at 1000 K for 60 min, (c) 120 min and (d) 180 min. These images support the very different kinetics for Cu encapsulation under MoS$_2$ when compared to Cu encapsulation under graphite. The "carpet" expands by material from the Cu islands diffusing through sputtered defects at the island base. Initially each island has small "carpet" areas around its perimeter; with time the separate "carpet" areas merge and at the end cover mesoscopic scale areas like in Fig. 5(d).

**Evidence of encapsulation from postdeposition annealing**

As discussed previously the presence of the "carpet" (and not compact, separate Cu islands encapsulated under graphite) suggests very different kinetics in the two cases. The "carpet" growth is fed by the islands on top which supply the material through sputtered defects at their base. This is consistent with the much lower island density nucleated on top and the much larger Cu islands on MoS$_2$ than on graphite. This scenario is confirmed from postdeposition annealing experiments shown in Fig. 5. After initial deposition at $T_{dep}$ = 1000 K, we annealed the sample in UHV at the same temperature for different time intervals. After each annealing step, the sample was cooled to room temperature and removed from the UHV system for SEM imaging.

A series of SEM images from the same location in Fig. 5 clearly shows the expansion of the intercalated Cu thin film at the expense of the 3D Cu clusters. Two clusters circled in red in Fig. 5(a) completely disappeared in Fig. 5(d) and resulted in a combined intercalated area of 25×25 μm$^2$. Another cluster circled in yellow had a corner missing after 180 min annealing and resulted in an intercalated annulus ~15 μm in diameter. At 180 min the right side of the imaged area has been fully intercalated with a larger fraction of the completion happening in the last 30 min. This indicates that some abrupt change specific to the nucleation of some of the islands can occur at random times. This change gives monotonic but super-linear expansion of the area of the "carpet". The sputtered defects under the island basis are expected to be filled with Cu adatoms. For example, for islands with several sputtered defects at their base of different sizes, the defects suddenly become "unclogged" because of the higher 1000 K temperature. This will provide more entry portals of material to move below, thus increasing the encapsulation rate.

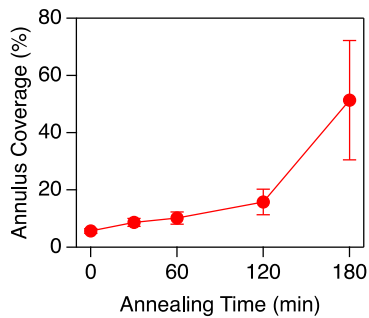

**Figure 6:** Total annular area (given as a fraction of the total area) of the series of images shown in Fig. 5 after postdeposition measured as a function of annealing time at 1000 K. As discussed in the text, the islands that most likely nucleate at defects provide the material to grow the "carpet" through defects under their base. As seen in Fig. 5(d), after the last annealing the right half of the area has been fully encapsulated and several islands at the top right have become extinct, which can account also for the acceleration of the growth rate. The initial expansion speed of the "carpet" is 13 μm$^2$/min prior to 120 min which increases to 58 μm$^2$/min for the remaining time.

The coverage of the total intercalated annulus area as a function of annealing time is plotted in Fig. 6 where 0 min indicates the initial surface after deposition. Only 3D islands are seen at the beginning but with time the annular area increases. In the analysis of postdeposition annealing growth, the fractional areas and its standard deviation were determined. For a given annealing step, several images were collected, the "carpet" areas were measured in each image; the average "carpet" area at a given time was determined. The annulus area was determined using the "flood" function in the WSXM software.

We assume a full coverage of intercalated Cu underneath the 3D clusters especially since this area should be completed first before the "carpet" expands outside the footprint of the measured island. From Fig. 6, one can see that the annulus coverage increases linearly during the first four annealing steps up to 120 min, and then growth is faster in the last step. This jump in annulus coverage could be a result of more entry portals (which were blocked by Cu adatoms) opening up at the high annealing temperature 1000 K.

**Evolution to non-equilibrium intercalated phase from the Cu pyramidal islands**

Different evidence has been described confirming that Cu intercalates below $MoS_2$ if deposited above 1000 K, based on a range of topographic and spectroscopic techniques. A simpler and more direct evidence of the intercalation is seen by the intercalated flat morphology, unique to $MoS_2$ so far: the formation of a homogeneous "carpet" that eventually covers the full area of the substrate if enough Cu is deposited. This observation by itself, very different from Cu encapsulation under graphite, is strong evidence that the deposited metal has been intercalated below and is not on top. As discussed in ref. [13], the deposited Cu in the form of islands at slightly lower temperature is in a non-equilibrium state because the islands have pyramidal shapes with the facet planes being the high index planes (311). At equilibrium, the island shapes should be a portion of truncated octahedra according to the Winterbottom construction, and the facetted planes should be predominantly low index planes. The (311) planes have higher energy than the low index planes (111) and (100). The observed non-equilibrium shapes are grown kinetically: as a balance between different processes controlling mass transport along and between the Cu layers forming the island, after the base of the island is first completed as an extended Cu(111) planar triangle. This triangular shape promotes the triangular symmetry of a 3-facetted pyramid. Annealing the pyramidal islands should drive them closer to their octahedral equilibrium shapes consisting primarily of (111) and (100) facets. These polyhedral shapes were calculated in [21]. Such an equilibration process and octahedral island shapes are not observed which excludes the possibility of atom spreading from the islands to wet $MoS_2$ and form the "carpet" on top of the substrate. It confirms that the "carpet" is not on top but is intercalated below the top $MoS_2$ TL.

A more general result applicable to graphene and all other 2D materials is the difficulty to wet them by metals after deposition on top. Metal growth on top of graphene was studied extensively and discussed in [22, 23] both exprerimenatlly and theoretically. All metals studied (Pb, Fe, Pt, Dy, Ru, Ag, Au, Gd, Eu) were found to grow 3D islands. Large 3D crystalline islands form at Liquid Nitrogen temperatures ($LN_2$ for weakly interacting metals like Pb) or 3D fractal islands (for strongly intercatingmetals like Gd, Dy). The growth of a larger number of metals was studied with DFT to confirm this general experimental result and to identify the driving force behind this. The metal cohesive energy is much larger than the metal-graphene interaction because the in-plane $sp^2$ bonds are very strong resulting in weak bonding normal to the surface. This competition applies to $MoS_2$ and all other 2D materials. This promotes the metal growth to be 3D. Since this is a thermodynamic force it becomes even stronger with increasing temperature. It is hard to see how Cu does not wet $MoS_2$ at 900 K but it wets at the higher temperature 1000 K. Wetting graphene has been an outstanding challenge in the community, especially in the context of developing uniform low resistance metal contacts. It was possible to wet graphene by growth manipulation to suppress transfer to higher layers in the crystalline islands, but this requires temperatures below LN$_2$ and stepwise coverage deposition [24].

## IV. Discussion

Studies of metal intercalation below bulk MoS$_2$ under UHV conditions are rather limited [25] because only recently the system has been examined as a primary 2D material with potential electronic and photonic applications. Our studies have shown a robust method to intercalate Cu under MoS$_2$ in a very unusual uniform morphology of a laterally extended layer, and not well-separated tall islands as in graphite. This is an indication that the expansion of the intercalated layer is through defects at the island base after deposition at 1000 K, and initially small patches of the "carpet" surround each island. With annealing, these patches expand until they merge so eventually the intercalated layer is spread over the whole substrate area. The intercalated layer has average thickness 5nm and is covered with a single MoS$_2$ TL. The expansion of the layer in the post-deposition annealing experiments shows that the area growth rate accelerates with time, most likely because defects under the island base become "unclogged" from Cu adatoms, so Cu feeds the carpet at higher rate.

One other study of Si intercalation under bulk MoS$_2$ was carried out with STS and XPS [25] at room temperature, with initially a high density of native defects. Uniform I-V spectra were recorded on a modified hills-and-valley morphology after Si deposition, signifying topographic surface re-arrangement due to mass transfer. After sputtering XPS measurements show an increase in the Si signal thus verifying that Si must be underneath; if Si was on top the Si signal should drop as Si is sputtered away.

Other limited metal intercalation work has been performed under single MoS$_2$ layer grown on different substrates. It is possible to grow high-quality single-layer MoS$_2$ on Au(111) with a single domain orientation[26], to selectively control the MoS$_2$ thickness, whether single or double layer [27, 28]; and on wafer-scale continuous single layer MoS$_2$ on sapphire characterized with azimuthal reflection high-energy electron diffraction [29]. Single layer MoS$_2$ grown on Au(111) was intercalated with Cesium (Cs) at 900 K under UHV conditions[30]. It was found that intercalation decouples MoS$_2$ from its substrate with some lattice expansion of MoS$_2$ where Cs bonds. The temperature range was determined for the reverse process of de-intercalation. In the previous, studies different techniques have been used to characterize the quality of the grown morphology of the single layer MoS$_2$ on the substrate used. For the two cases of intercalation mentioned (Si, Cs), aspects of both the structure and electronic properties of the intercalated systems have been investigated.

The emphasis in the current work was to identify a different intercalation mechanism than on graphite, the key controlling processes and the type of mass transport controlling atom transfer. The results of the current experiments show that a flat intercalated phase is grown and that defects at the base of the Cu islands provide the portals for Cu to move below. Such information can identify the key kinetic processes, so the metal encapsulation becomes more predictive. This information can be useful in other metal intercalation studies under bulk or single layer MoS$_2$; especially to build theoretical models that describe how the metal atoms move below and the energetics of each intercalated phase. As was demonstrated for graphene intercalation in the absence of intentionally introduced defects, other locations of low symmetry (steps, domain boundaries, anti-phase boundaries, etc.) can also be entry portals for metal atoms to move below the 2D material [31].

## V. Conclusion

Intercalation of Cu underneath bulk MoS$_2$ through sputtered defects was demonstrated with the use of different structural and spectroscopic techniques. SEM was used to probe the morphological changes after deposition of Cu above 1000 K and transfer of Cu to lower MoS$_2$ galleries. Below this temperature only islands form on top, but above it Cu moves below and expands laterally producing a mesoscopic scale intercalated layer of ~5 nm average thickness. The transfer below is through defects at the base of the islands. EDS shows that in this layer there is strong Cu signal. Differences in the shape of different XPS Cu peaks after deposition at 900 K (with only islands a on top) when compared to deposition at 1000 K (with Cu layer below MoS$_2$) confirm the intercalation. In the latter case, inelastic scattering through MoS$_2$ results in larger fraction of ejected photoelectrons with lower energy, which generates asymmetry in the measured XPS shapes. This is further confirmed by comparing the larger attenuation of Cu 2$p$ than Auger photoelectrons because of lower energy exiting MoS$_2$. These experiments produce a uniform intercalated layer with more homogeneous electronic properties (and not separate individual islands of larger height as under graphite). More importantly, they demonstrate that the general method developed for graphite intercalation through sputtered defects can be generalized to other more complex 2D materials.

## Acknowledgments

This work was supported by the U.S. Department of Energy (DOE), Office of Science, Basic Energy Sciences (BES), Materials Science and Engineering Division. JWE was supported by the U.S. DOE BES Division of Chemical Science, eosciences, and Biological Sciences. The research was performed at Ames Laboratory, which is operated for the U.S. DOE by Iowa State University under contract # DE-AC02-07CH11358.

```
```